# 3.3 Gigahertz Clocked Quantum Key Distribution System

Karen J. Gordon (1), Veronica Fernandez (1), Robert J. Collins (1), Ivan Rech (2),
Sergio D. Cova (2), Paul D. Townsend (3), and Gerald S. Buller (1)

1: David Brewster Building, School of Engineering and Physical Sciences, Heriot-Watt University, Riccarton,
Edinburgh, EH14 4AS, UK.  k.j.gordon@hw.ac.uk
2: Dipartimento Elettronica e Informazione, Politecnico di Milano, 20133, Milano, Italia.
3: Tyndall National Institute and Department of Physics, University College Cork, Cork, Ireland.

**Abstract** A fibre-based quantum key distribution system operating up to a clock frequency of 3.3GHz is presented. The system demonstrates significantly increased key exchange rate potential and operates at a wavelength of 850nm.

**Introduction**
Quantum key distribution (QKD) enables a verifiably secret key to be shared between two parties Alice and Bob, with security guaranteed by the laws of quantum mechanics [1,2].

We present an improved QKD test system operating at clock rates up to 3.3GHz by using a specially adapted commercially-available silicon single-photon counting module and improved laser driving electronics. The QKD system [3] implements the B92 protocol [4] using polarisation encoding, and was designed to establish keys at high key exchange rates over relatively short transmission distances of standard telecommunications fibre (<20km). The use of an enhanced silicon detector has improved the fibre-based QKD system in terms of transmission distance and quantum bit error rate (QBER).

**Description of the QKD System**
This QKD system, shown in figure 1, operates at a wavelength of 850nm, which enabled the use of technologically advanced silicon single photon avalanche diodes SPADs and the possibility of high key exchange rates of the order of ~1Mcount/s to be achieved. At this wavelength, the fibre loss is ~2.5dBkm$^{-1}$ limiting the transmission range of the QKD system to less than 20km.

Two polarised, attenuated vertical-cavity surface-emitting lasers (VCSELs) were employed, in order to provide a weak coherent source of photons. The VCSELs were capable of operating at frequencies >>1GHz, and were temperature-tuned to have identical peak wavelengths, thereby avoiding spectral interrogation by an eavesdropper.

The transmitter (Alice) and receiver (Bob) were constructed from 850nm single mode fibre; however the transmission medium between Alice and Bob was standard 9μm diameter core telecommunications fibre. Although standard telecoms fibre supports several spatial modes at 850nm, stable single mode propagation was achieved in the experiment by fusion splicing the two types of fibre concentrically together to obtain a mode-selective launch [5]. Under these conditions the measured fibre loss was ~2.2dBkm$^{-1}$, which was close to the anticipated value for 850nm operation..

The linearly polarised light from each VCSEL was launched into 850nm single mode fibre, and the relative polarisation angle set to 45° using fibre polarisation controllers (PCs). The silicon SPADs detected the photons filtered by Bob's polarising beam splitters (PBSs). Each 850nm photon is "time stamped" using an optical synchronisation pulse, of wavelength 1.3μm, generated by a distributed feedback laser (DFB), and detected by a germanium avalanche photodiode (APD) biased below breakdown.

Recent improvements in the QKD system performance by increasing the clock frequency from the original 1GHz rate [3] to the 3.3GHz value achieved here have been feasible due to the introduction of an improved SPAD module and improved VCSEL driving electronics [6].

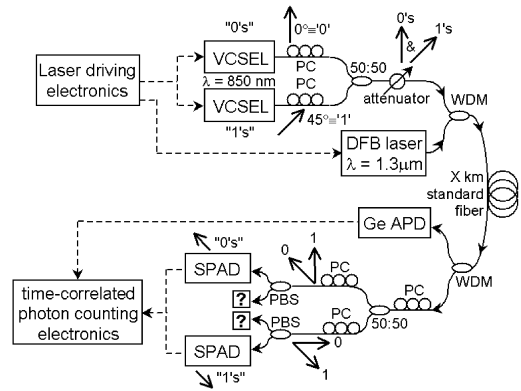

Figure 1. Schematic diagram of the quantum key distribution experiment. PBS: Polarising beam splitter. PC: Polarisation controller. WDM: Wavelength division multiplexer. APD: Avalanche photodiode. VCSEL: Vertical-cavity surface-emitting laser. SPAD: Single photon avalanche diode. DFB: Distributed feedback laser.

**The Enhanced SPAD Module**
The SPAD module (a standard Perkin Elmer SPCM-AQR photon detector module) was improved in terms of photon timing performance by inserting an additional circuit card using the technique described in [7,8]. This modification was reversible and the additional circuit card could be removed with no detrimental effects to the original circuit card.

A pulse pick-off linear network was connected to the SPAD terminal, which was biased at a high-voltage (about 400V). The network was specifically designed to extract a short pulse signal with fast rise, practically coincident with the rise of the avalanche current. A fast discriminator with very low sensing threshold was then employed for sensing the onset of this pulse. Therefore, the avalanche current can be sensed at an initial stage of its build-up, when still confined in a small area of the detector. Thus, the time information obtained was not affected by the statistical fluctuations that characterize the propagation of the current over the full area of the detector [9]. Hence, the jitter in the measured arrival time of the photon was minimized, as shown in figure 2.

distances, ~50,000bit/s at 6.55km and ~6000bit/s at 11km.

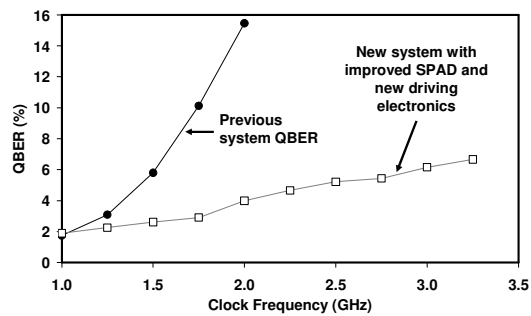

Figure 3. Graph of improvement in QBER versus QKD system clock frequency for a fixed fibre distance of 6.55km.

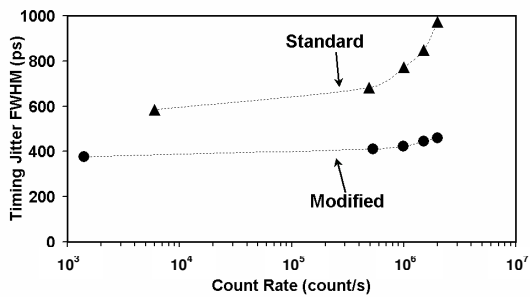

Figure 2. Timing jitter full width at half maximum of the standard SPCM SPAD and the SPCM with modified output circuitry.

At low counting rates the original module prior to modification had a full width at half maximum (FWHM) jitter of ~570ps, but with the modification the module had an improved FWHM jitter of ~370ps. The additional circuit card has an added advantage in that its performance was more stable at high pulse counting rates (typically above 0.5Mcount/s).

Temporal broadening of the single-photon detector response has been shown to limit the performance of the QKD system [3] since at clock frequencies between 1 and 3.3GHz and at shorter fibre distances the detected count rate can be up to 4Mcount/s.

**Experimental Results**
We will present experimental data over a range of system clock frequencies in terms of quantum bit error rate (QBER) and the final sifted key exchange rate after error correction and privacy amplification. Figure 3 shows a graph of the improvement in QBER versus the system clock frequency of the QKD system, at a fixed distance of 6.55km between Alice and Bob. Key exchange rates after error correction and privacy amplification have been estimated to be of the order of ~1Mbit/s, for very short transmission

**Conclusions**
The temporal response of a commercially available single-photon counting module has been significantly improved via a relatively low-cost modification, consisting of the addition of a single dedicated circuit board. The modified detector has been shown to offer important benefits when applied in a QKD system operating at clock rates in excess of 1GHz. Specifically the QKD system has been improved in terms of increased workable clock frequency range from 1GHz to 3.3GHz, with the results at higher frequencies demonstrating the potential for increased transmission distance.